\documentclass[aps, twocolumn,floats, showpacs]{revtex4}
\usepackage{amsmath,amssymb,graphicx}
\usepackage{epsfig}
\usepackage{graphicx}
\usepackage{enumerate}

\newcommand{\beq}{\begin{equation}}
\newcommand{\eeq}{\end{equation}}

\newcommand{\bea}{\begin{eqnarray}}
\newcommand{\eea}{\end{eqnarray}}

\begin{document}

\title{\footnotetext{APS copyright, to be printed in Phys. Rev. Lett.}
Stochastic pumping of heat: Approaching the Carnot efficiency}
\author{Dvira Segal}
\affiliation{Chemical Physics Theory Group, Department of
Chemistry, University of Toronto, 80 St. George Street, Toronto,
Ontario M5S 3H6, Canada}

\date{\today}
\begin{abstract}

Random noise can generate a unidirectional heat current across 
asymmetric nano objects in the absence (or against) a temperature gradient.
We present a minimal model for a molecular-level stochastic heat
pump that may operate arbitrarily close to the Carnot efficiency.
The model consists a fluctuating molecular unit
coupled to two solids characterized by distinct phonon spectral properties.
Heat pumping persists for 
a broad range of system and bath parameters. Furthermore, by filtering the reservoirs' phonons
the pump efficiency can approach the Carnot limit.

\end{abstract}

\maketitle


Heat flows spontaneously from objects at high temperature to objects
of low temperature. Directing heat against a temperature gradient
requires the application of an external perturbation or the design of a special system-bath initial condition.
A generic nano-scale heat pump model includes a molecular element bridging two solids.
An external force modulates the subsystem (molecular) energetics,
leading to a net exchange of heat between the baths against, or in the absence, of a temperature gradient.
A spatial asymmetry should be built into the system in order to define a preferential directionality.
This process is closely related to the ratchet effect where a particle current is catalyzed
in the absence of a dc voltage drop due to the asymmetry in the
spatial potential and the influence of time dependent forces
\cite{Reimann}.

Recent studies have carefully analyzed the operation principles of
prototype classical \cite{Alla3,Komatsu,Dhar} and quantum
\cite{Geva,Linke, Nori,Alla,Michel,Pump,Kohler}
thermal machines, seeking to optimize performance. Typically, in these
schemes a {\it carefully-shaped} external force  periodically
modulates the levels of the nano object, leading to the pumping operation. 
In electronic systems a Brownian electron refrigerator based on selective 
tunneling through a metal-insulator-superconductor 
junction was recently proposed \cite{Pekola}. 
Fluctuating electric fields can also do chemical work, driving ions
against an electrochemical potential difference \cite{Astumian}.

In this paper we demonstrate facile pumping of heat in the form of
vibrational energy in response to {\it random noise}.
The following generic effect is discussed:
Heat can be pumped between distinct solids 
when the bridging object suffers stochastic fluctuations of its energies, 
given that the spectral density of at least one solid
strongly varies within the noise spectral window.
Furthermore, using reservoirs with suitable energy filters, (Einstein solids)
one can construct a heat pump that works close to the Carnot limit.
Thus, we demonstrate that not only can one utilize random noise and pump heat against a bias ("Maxwell demon"),
the efficiency of this machine may approach the maximal value.




Our minimal model consists a Kubo oscillator \cite{Kubo} bilinearly coupled to two macroscopic
reservoirs.
The model could be realized by adsorbing a molecule 
on a dielectric surface, blocking charge transfer, while allowing
for heat exchange between the modules. The total Hamiltonian
includes the following contributions,
$ H=H_0(t)+H_B+V_L+V_R$. 
%
$H_0$ includes the isolated subsystem, a Kubo oscillator,
\bea H_0(t)=\sum_{n=0,1..\infty}E_n(t)|n\rangle \langle n|; \,\,\,\,
E_n(t)=\left[ E_n^{(0)}+\epsilon_n(t)\right]. \label{eq:H0}
 \eea
$E_n^{(0)}=nB_0$ is the static energy of level $|n\rangle$,
$\epsilon_n(t)$ is a time dependent stochastic contribution. We
assume that the average over  stationary fluctuations vanishes, $\langle
\epsilon(t)\rangle_{\epsilon}$=0, while higher  moments survive.
This local oscillator is coupled to two thermal baths of temperatures 
$T_{\nu}=1/\beta_{\nu}$  ($k_B\equiv
1$), represented by sets of independent harmonic oscillators ($\hbar\equiv 1$)
\bea H_{B}&=&\sum_{\nu,k} \omega_k b_{\nu,k}^{\dagger}b_{\nu,k}.
\label{eq:HB} \eea
$b_{\nu,k}^{\dagger}$, $b_{\nu,k}$ are the bosonic creation and
annihilation operators respectively for the mode $k$ of  bath $\nu=L,R$.
The interaction between the
subsystem and the bath is taken to be bilinear (displacement-displacement) \cite{NDR},
\bea V_{\nu}&=&  F_{\nu} \sum_{n=1,2..\infty} c_{n}|n\rangle \langle n-1| +c.c.
\nonumber\\
F_{\nu}&=&\sum_{k}
\lambda_{\nu,k}(b_{\nu,k}^\dagger+b_{\nu,k}), \label{eq:V}
\eea
where $c_{n}=\sqrt{n}$, and the system-bath interaction is
characterized by the spectral function
\bea g_{\nu}(\omega)=2 \pi \sum_{\nu,k}\lambda_{\nu,k}^2
\delta(\omega-\omega_k). \label{eq:g} \eea
%
%
In what follows we simplify the discussion and consider a two-level
system (TLS) ($n$=0,1) for the local Hamiltonian,
\bea H_0^{TLS}(t)&=& B(t) \sigma_z/2 ; \,\,\,\,\,
B(t)=\left[B_0 + \epsilon(t) \right]
\nonumber\\
V_{\nu}^{TLS}&=&F_{\nu}\sigma_x. \label{eq:HTLS} \eea
Here $\sigma_x=|1\rangle \langle 0| + |0\rangle \langle 1|$,
$\sigma_z=|1\rangle \langle 1| - |0\rangle \langle 0|$,
$B_0=E_1^{(0)}-E_0^{(0)}$, and
$\epsilon(t)$ is the stochastic modulation of this energy gap. We
sometimes refer to the truncated harmonic mode as a "spin".
In the weak coupling limit
the TLS dynamics can be trivially carried back to the harmonic limit
($n=0,1,... \infty$), as we discuss below.
%


{\it Stochastically averaged master equation.}
 Closed kinetic equations for the $|n\rangle$ state population
can be obtained in the weak system-bath coupling limit
\cite{Goychuk}. Briefly, starting with the
quantum Liouville equation, this involves (i) studying the quantum dynamics within 
second order perturbation theory, (ii) going into the markovian limit, assuming short correlation time of bath
fluctuations, and (ii) averaging over the stochastic process under
the decoupling approximation,
\bea \langle e^{i\int_0^x \epsilon(t')dt'}
P_{n}(x)\rangle_{\epsilon} \sim \langle e^{i\int_0^x
\epsilon(t')dt'}\rangle_{\epsilon} \langle P_n(x)\rangle_{\epsilon}.
\label{eq:decomp} \eea
Here $P_n$ ($n=0,1$) denotes the TLS population, and the average is
performed over energy fluctuations. The decomposition
(\ref{eq:decomp}) relays on the separation of timescales: Energy
fluctuations are assumed to be  fast in comparison to the
characteristic subsystem relaxation time.
The three assumptions (i)-(iii) result in the following 
equations  $\langle  P_1(t)  \rangle_{\epsilon} +
\langle  P_0(t) \rangle_{\epsilon}=1$,
\bea  \langle \dot P_1(t)  \rangle_{\epsilon}
&&=-\left(k^L_{1\rightarrow 0} +k^R_{1\rightarrow 0}\right)
 \langle P_1(t) \rangle_{\epsilon}
\nonumber\\ &&+\left( k_{0\rightarrow 1}^L +k_{0\rightarrow
1}^R\right) \langle P_0(t) \rangle_{\epsilon},
 \label{eq:dyn} \eea
with the activation and relaxation rates 
\bea &&k_{0 \rightarrow 1}^{\nu}= \int_{-\infty} ^\infty d\omega
g_{\nu}(\omega) N_{\nu}(\omega) I(B_0-\omega),
\nonumber\\
&&k_{1 \rightarrow 0}^{\nu}= \int_{-\infty} ^\infty d\omega
g_{\nu}(\omega)[N_{\nu}(\omega)+1]I(B_0-\omega).
\label{eq:krate} \eea
Here $g_{\nu}(\omega)$  is the phonon spectral function
defined in (\ref{eq:g}), 
extended to negative values $g_{\nu}(\omega)=-g_{\nu}(-\omega)$, 
$N_{\nu}(\omega)=[e^{\beta_{\nu}\omega}-1 ]^{-1}$
is the Bose-Einstein occupation factor
and 
\bea I(\omega)=\frac{1}{2\pi}\int_{-\infty} ^{\infty} dt e^{i\omega
t} \langle e^{i\int_0^t \epsilon(t') dt'}\rangle_{\epsilon},
\label{eq:spec} \eea
is the spectral lineshape of the Kubo oscillator.
Assuming a Gaussian process, going to the
fast modulation limit $\langle \epsilon(t_1)\epsilon(t_2)\rangle
=2\gamma\delta(t_1-t_2)$,  the noise spectral
lineshape can be approximated by a Lorentzian function
\bea I(\omega)=\frac{\gamma/\pi}{\omega^2+\gamma^2} \label{eq:Loren}
\eea
of width $\gamma$. 
For $\gamma\rightarrow 0$,  $I(\omega)=\delta(\omega)$, and
the field-free vibrational relaxation rates \cite{Berne} are
retrieved from Eq. (\ref{eq:krate}): The rates are evaluated at the
bare TLS frequency, $k^{\nu}_{0\rightarrow
1}=g_{\nu}(B_0)N_{\nu}(B_0)$, $k_{1\rightarrow
0}^{\nu}=g_{\nu}(B_0)[N_{\nu}(B_0)+1]$, 
obeying the detailed balance condition.
In contrast, for a general driving this
condition is violated, as the time dependent field drives the
subsystem out of thermal equilibrium with the bath. 

The master equation (\ref{eq:dyn})
can be trivially generalized in the weak coupling regime to describe a harmonic local
mode [Eqs. (\ref{eq:H0})-(\ref{eq:V})] under a stochastic field,
%
\bea &&\langle \dot P_n\rangle_{\epsilon}=
-\sum_{\nu}[(n+1)k^{\nu}_{n \rightarrow n+1}+nk^{\nu}_{n \rightarrow
n-1}]\langle P_n\rangle_{\epsilon}
\nonumber\\
&&+ (n+1)\sum_{\nu}k^{\nu}_{n+1\rightarrow n} \langle
P_{n+1}\rangle_{\epsilon}+ n\sum_{\nu}k^{\nu}_{n-1\rightarrow n}
\langle P_{n-1}\rangle_{\epsilon}.
\nonumber\\
\label{eq:dynH} \eea
Note that for harmonic systems the microscopic rates are independent of the index $n$, 
thus we only identify activation and relaxation processes using  (\ref{eq:krate}). 

\begin{figure}[htbp]
\hspace{2mm} \rotatebox{270}{\hbox{\epsfxsize=30mm
\epsffile{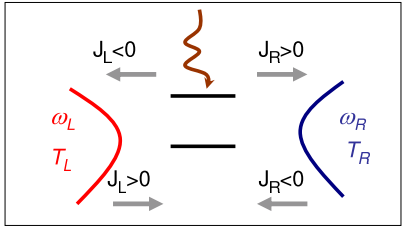} }}
\caption{Scheme of the stochastic pump.
A molecular element under an external perturbation (curly line)
is coupled to two solids of different spectral properties.
The arrows display the sign notation.
}
 \label{Fig0}
\end{figure}

{\it Heat current.} We derive an analytical expression for the
heat flux in the presence of stochastic energy modulations. We begin
with the TLS model (\ref{eq:HTLS}), then extend the results to
the harmonic case. 
An expression for the energy flux operator,
applicable in both stationary and time dependent
situations, was given in \cite{Current},
$\hat J_R=\frac{i}{2} \left[ (H_0^{TLS}-H_B), V_R^{TLS} \right]$.
%
The current is defined positive when flowing from  $L$ to  $R$,
see Fig. \ref{Fig0}. 
We substitute Eqs. (\ref{eq:HB}) and
(\ref{eq:HTLS}) into this expression and obtain 
\bea \hat J_R(t)=- \frac{1}{2}\big[ \sigma_x G_R +B(t) \sigma_y F_R
\big], \label{eq:curr}
 \eea
with $\sigma_y=-i|1\rangle\langle0| + i|0\rangle \langle 1|$,
$F_{\nu}=\sum_{k}\lambda_{\nu,k}(b_{\nu,k}^{\dagger}+b_{\nu,k})$,
$G_{\nu}=i\sum_{k}\lambda_{\nu,k} \omega_k
(b_{\nu,k}^{\dagger}-b_{\nu,k})$.
%
The expectation value of the current is given by $J={\rm Tr}\{\rho
\hat J\}$, where $\rho$ is the total density matrix, and the trace
is performed over the thermal bath and the subsystem degrees of
freedom. A further averaging over the TLS energy fluctuations is
necessary. Since in steady-state the condition $\langle {\rm Tr}\{
\partial V_R/\partial t\} \rangle_{\epsilon} =0$ translates to $\langle {\rm Tr }
\{G_R \sigma_x\}\rangle_{\epsilon}= \langle {\rm Tr} \{B(t) \sigma_y F_R
\}\rangle_{\epsilon}$, the stationary  noise-averaged current can
be calculated by studying  the first term in (\ref{eq:curr}),
%
$\langle J_R\rangle_{\epsilon}=
-\langle{\rm
 Tr}_B\{\rho_{1,0}G_R+\rho_{0,1}G_R \}\rangle_{\epsilon}$.
%
Here $\rho_{i,j}$ are the matrix elements of the density matrix, and
${\rm Tr}_B$ denotes a trace over the bath states.
Next, using a second order expansion of $\rho_{1,0}$, adopting
the same set of approximations  employed above for deriving
(\ref{eq:dyn}), we acquire the {\it steady-state} heat current expression
\bea \langle J_R\rangle _{\epsilon}= \langle P_1\rangle_{\epsilon}
f_{1\rightarrow 0}^R -\langle P_0\rangle_{\epsilon}f^R_{0\rightarrow
1}. \label{eq:currE} \eea
The noise-averaged population is obtained by solving 
(\ref{eq:dyn}) in steady-state, taking $\langle \dot P_n\rangle
_{\epsilon}=0$.
The energy absorption and dissipation rates (Energy/time) are
given by
\bea f^{\nu}_{0\rightarrow 1}&=&\int_{-\infty}^{\infty} d\omega
\omega g_{\nu}(\omega)I(B_0-\omega)N_{\nu}(\omega);
\nonumber\\
f^{\nu}_{1\rightarrow 0}&=& \int_{-\infty}^{\infty} d\omega \omega
g_{\nu}(\omega)I(B_0-\omega)[N_{\nu}(\omega)+1]. \label{eq:frate}
\eea
%

Note that in the absence of fluctuations the rates satisfy the relation
$f_{n\rightarrow n\pm 1}^{\nu}=B_0 k^{\nu}_{n\rightarrow n\pm 1}$
\cite{Berne}, while in the presence of stochastic force we
cannot trivially connect the energy transition rates
(\ref{eq:frate}) to the population relaxation rates (\ref{eq:krate}).
In the weak coupling limit Eq. (\ref{eq:currE}) can be easily
generalized to the harmonic (Kubo) case,
%
$\langle J_R\rangle _{\epsilon}=\sum_{n=1}^{\infty} n[\langle P_n\rangle_{\epsilon}
f^R_{n\rightarrow n-1} -\langle P_{n-1}\rangle_{\epsilon}f^{R}_{n-1\rightarrow n}]$,
%
with the population  obtained by solving (\ref{eq:dynH}) in steady-state.
%


{\it Pumping mechanism.} The pumping mechanism relays
on two conditions: (i) The spectral function (of at least one bath)
should strongly vary in the frequency window of the noise lineshape.
(ii) The phonon reservoirs should have distinct spectral properties.
We assume that $\omega_R \lesssim B_0 <\omega_L$;
 $\omega_{\nu}$ is the $\nu$ contact cutoff vibrational frequency. 

Due to the fast stochastic modulation, the spacing $B(t)$ can be abruptly
reduced to values below $\omega_R$. Since these modulations are faster
that the subsystem relaxation time, the  TLS effective temperature
becomes significantly low, even lower than $T_{R}$. The TLS then
absorbs energy from both baths. In contrast, fluctuations
that largely increase the gap lead to a high effective internal
temperature, while practically disconnecting the $R$ bath from the system.
If the internal temperature is higher than $T_L$,  energy will
dissipate from the Kubo mode to the left bath. Thus,
in this asymmetric setup net energy can be injected from the $R$ bath to the $L$ bath even for $T_L>T_R$.
Note that this principle is independent of the specific
form assumed for the noise spectral function, the molecular properties, and the details of the
baths.

This mechanism is reminiscent to the idea proposed in \cite{Pump},
with two crucial distinctions: (i) Here random perturbations catalyze (maybe even undesirably)
heat flow, rather that carefully shaped pulses. 
(ii) The pump operates in the fast modulation regime, rather than adiabatically/quasiadibatically,
leading to the Carnot efficiency under proper conditions.



{\it An exactly solvable model. }
We employ next the  Einstein model for the phonon spectral functions,
$g_{\nu}(\omega)=\xi_{\nu}\delta(\omega-\omega_{\nu})$, $\xi_{\nu}=2\pi\lambda_{\nu}^2$ from (\ref{eq:g}).
Practically, this scenario may be realized by filtering the contact's phonon modes \cite{Filter}.
Assuming that $I(B_0+\omega_{\nu})\ll I(B_0-\omega_{\nu})$ \cite{commentN},
the transition rates (\ref{eq:krate}) and (\ref{eq:frate}) reduce to
\bea && k_{0\rightarrow
1}^{\nu}=\xi_{\nu}I(B_0-\omega_{\nu})N_{\nu}(\omega_{\nu});
\nonumber\\
&&k_{1\rightarrow
0}^{\nu}=\xi_{\nu}I(B_0-\omega_{\nu})\left[N_{\nu}(\omega_{\nu})+1\right];
\nonumber\\
&& f_{0\rightarrow
1}^{\nu}=\omega_{\nu}\xi_{\nu}I(B_0-\omega_{\nu})N_{\nu}(\omega_{\nu});
\nonumber\\
&&f_{1\rightarrow
0}^{\nu}=\omega_{\nu}\xi_{\nu}I(B_0-\omega_{\nu})\left[N_{\nu}(\omega_{\nu})+1\right],
 \eea
and the energy current at the $\nu$ contact becomes \cite{NDR} 
\bea \langle J_{\nu}\rangle_{\epsilon}=\omega_{\nu} {\mathcal T}
\left[N_L(\omega_L)-N_R(\omega_R)\right]
\eea
where ${\mathcal T}= \frac{\gamma_L(\omega_L)\gamma_R(\omega_R)}{\gamma_L(\omega_L) +
\gamma_R(\omega_R)}$ for the  Hamiltonian (\ref{eq:H0})-(\ref{eq:V}) and
${\mathcal T =}\frac{\gamma_L(\omega_L)\gamma_R(\omega_R)}{ \sum_{\nu}\gamma_{\nu}(\omega_{\nu})[1+2N_{\nu}(\omega_\nu)]}  $
for the TLS model (\ref{eq:HTLS}).
Here $\gamma_{\nu}(\omega_{\nu})\equiv\xi_{\nu}I(B_0-\omega_{\nu})$.
In both cases the $R$ contact is cooled down when $N_R(\omega_R)>N_L(\omega_L)$,
i.e. for $\beta_L\omega_L> \beta_R \omega_R$. Defining $\delta T=T_L-T_R>0$,
$\delta \omega=\omega_L-\omega_R>0$, the pumping condition translates to
\bea
\frac{\delta T}{T_R} <\frac{\delta \omega}{\omega_R}.
\label{eq:ineq}
\eea
%
For reversible systems the net current is zero, and an equality is attained, $\delta T/T_R = \delta \omega /\omega_R$.
Note that for $\delta T=0$ a unidirectional current develops merely based on the asymmetry $\omega_L>\omega_R$.
We calculate next the machine efficiency.
In steady-state the work over time done by the external force is given by the difference
$\langle W\rangle_{\epsilon}=\langle J_R\rangle_{\epsilon}-\langle J_L\rangle_{\epsilon}$,
\bea \langle W\rangle_{\epsilon}= (\omega_{R}-\omega_L) {\mathcal T}
\left[N_L(\omega_L)-N_R(\omega_R)\right].
\eea
The averaged cooling efficiency $\eta \equiv -\langle J_R\rangle_{\epsilon}/\langle W\rangle_{\epsilon}$ (the minus emerges due to
the sign notation) then becomes
\bea
\eta= \frac{\omega_R}{\delta \omega} < \frac{T_R}{\delta T}\equiv \eta_{max},
\eea
where the inequality was obtained using (\ref{eq:ineq}). Thus, the maximal efficiency of our cooling device
is exactly given  by the Carnot efficiency, reached when the device works reversibly.
This result is independent of system properties (spin vs. harmonic mode) and the details of the noise lineshape.
Finally, we verify that the average entropy production per unit time is always positive for an irreversible device,
$\langle \sigma\rangle_{\epsilon}={\mathcal T} \left(\frac{\omega_L}{T_L}-\frac{\omega_R}{T_R}\right)\left[N_R(\omega_R)-N_L(\omega_L)\right]$.

\begin{figure}[htbp]
\hspace{2mm} {\hbox{\epsfxsize=70mm \epsffile{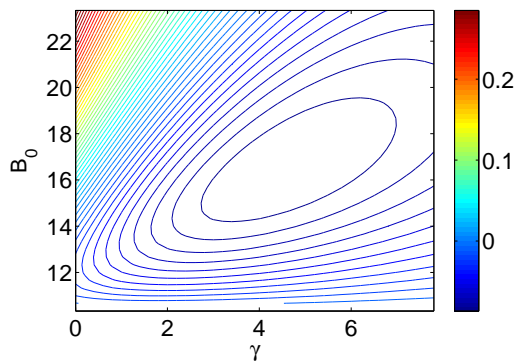}}}
\caption{Color-map of the heat flux at the $R$ contact, revealing a
pumping regime (central oval); $\omega_L=200$, $\omega_R$=3,
$T_L=T_R=25$. For these parameters $\langle J_L\rangle_{\epsilon}$ (not plotted) is negative.
} \label{Fig1}
\end{figure}
\begin{figure}[htbp]
\hspace{2mm} {\hbox{\epsfxsize=65mm \epsffile{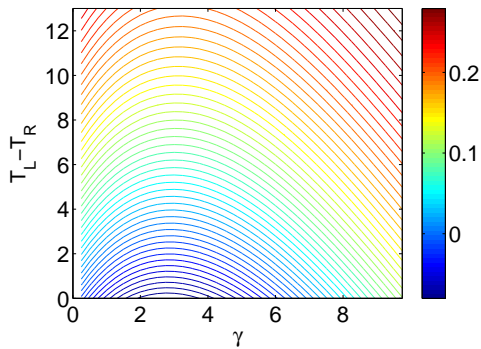}}}
\caption{A color-map of the heat flux at the $R$ contact 
where $\langle J_R\rangle_{\epsilon}$ is negative for $\gamma<8$ and
$T_L-T_R\lesssim 4$.
$\omega_L=200$, $\omega_R$=3, $B_0=15$, $T_R=25$.
}
\label{Fig2}
\end{figure}

{\it Numerical results.}
We exemplify the operation of the heat pump
by simulating  Eq. (\ref{eq:currE}) with the stationary population
determined by solving (\ref{eq:dyn}) in steady-state. 
We adopt the Lorentzian lineshape (\ref{eq:Loren}),
and assume that both solids are
characterized by the ohmic spectral function  $g_{\nu}(\omega)=A_{\nu}\omega e^{-\omega/\omega_{\nu}}$. We take
$\omega_L\sim 100$ and $\omega_R\sim B_0\sim 10$, $A_{L,R}=1$.
The calculations are performed in dimensionless units. If the energy parameters
are given in meV, the current divided by the factor
$\hbar=4.14$ (meV ps) yields fluxes in units of meV/ps.

First we assume that $T_L=T_R$,
 and demonstrate a unidirectional heat flow from $R$ to $L$.
Fig. \ref{Fig1} presents a color-map of the heat current at the $R$ side.
The flux at the central oval ($\gamma>2$; $12<B_0<20$) is negative,
i.e. the $R$ bath is cooled down. We can explain
this behavior as follows. For the ohmic function
with $\omega_R=3$ the $R$ bath is practically disconnected from
the TLS at energies of the order $B(t)\gtrsim15$. 
Therefore, for $B_0\sim 15$ fluctuations that reduce the gap
lead to an overlap with the $R$ phonon spectral function, and the
TLS can accept energy from this contact.
Fluctuations that significantly increase the gap ($B(t)>15$) imply
a complete isolating of the $R$ contact, 
as its spectral function does not overlap with the molecular vibrational frequency.
Energy is then solely exchanged between the (hot) TLS and the $L$
solid, resulting  in a net heat flow.


We further  demonstrate pumping of heat against a temperature gradient. 
Fig. \ref{Fig2} shows that $\langle J_R\rangle_{\epsilon}$ is negative for $\gamma<8$ and
$T_L-T_R\lesssim 4$. We verified that the total entropy of the
system  is  increasing.
Note that the pumped current is rather small, yet the object of this
calculation is to demonstrate the ubiquity of the pumping effect,
rather than to develop optimization schemes.

In summary, nano-objects may pump heat effectively  
by exploiting random fluctuations, given that the (distinct) solids are characterized
by frequency dependent spectral functions. 
Our work reverses standard perceptions of microscopic heat pumps:
(i) Directing heat at $\delta T=0$,
and pumping heat at finite $\delta T$ are facile processes, taking place in response to a general noise lineshape. 
(ii) Instead of carefully shaping the external pulse operating on the subsystem,
simple bath "shaping" (mode filtering) may lead to  maximal performance.

Temperature reduction of nano-level objects is important, e.g. 
 for cooling electronics devices, and 
for controlling chemical reactions and molecular dynamics. 
The pumping mechanism described here is pertinent to other systems, as  one could realize
an analogous exciton heat pump  \cite{radiation}.
The effect might be observed in various systems suffering random
noise, for example, in nanomechanical resonators. 
Considering a double clamped nano beam, the natural frequency could
be mechanically modulated due to the beam's
mass variation resulting from adsorption-desorption processes \cite{Roukes-mass}.
The resonator frequency could be also electrostatically tuned by
(stochastically) varying the bias applied to a gate electrode
\cite{McEuen}.

{\bf Acknowledgments} The work was supported by NSERC and by the University of Toronto Start-up Fund.





\begin{thebibliography}{32}
\bibitem{commentN}
The random field does not invert the levels, i.e. $E_1(t)>E_0(t)$.
\end{thebibliography}
\end{document}